\documentstyle[editedvolume]{crckapb} 

\def\abstract{\vspace*{27pt}{ABSTRACT}\par\relax}
\def\acknow{\par ACKNOWLEDGMENTS\par}

\newcommand{\stt}{\small\tt}

\begin{opening}
\title{MASSIVE ACCRETION DISKS}

\author{N.Z. SCOVILLE}
\institute{California Institute of Technology\\
           Astronomy Department 105-24\\
           Pasadena, CA 91125, USA}

\end{opening}

\runningtitle{MASSIVE ACCRETION DISKS}

\begin{document}

\section{Abstract}

Recent high resolution near infrared (HST-NICMOS) and mm-interferometric 
imaging have revealed dense gas and dust accretion disks in nearby ultra-luminous galactic
nuclei. 
In the best studied ultraluminous
IR galaxy, Arp 220, the 2$\mu$m imaging shows dust disks in both of the 
merging galactic nuclei and mm-CO line imaging indicates molecular gas
masses $\sim$10$^9$M$_{\odot}$ for each disk. The two gas disks in Arp 220 are counterrotating
and their dynamical masses are $\sim$ 2x10$^9$M$_{\odot}$, that is, only slightly larger
than the gas masses. These disks have radii $\sim$100 pc and thickness 10-50 pc. The high brightness temperatures of the CO lines indicate
that the gas in the disks has area filling factors $\sim$25-50$\%$ and mean densities
of $\geq 10^4$ cm$^{-3}$. Within these nuclear disks, the rate of massive star formation is undoubtedly prodigious and, given the high viscosity of the gas, there will also be high radial accretion rates, perhaps $\geq$10 M$_{\odot}$ yr $^{-1}$.  If this inflow persists to very small radii, it is enough to feed even the highest luminosity AGNs.

\section{Introduction}

In the luminous infrared galaxies, very large masses of interstellar
matter have been concentrated in the galactic nuclei at radii less than 300 pc
as a result of galactic merging.  Due to its relatively large volume-filling factor and ability to cool rapidly, the ISM is much more dissipative than the stars.  This dissipation of rotational energy 
and angular momentum will concentrate the ISM in the merging galactic nucleus.  The resultant high density of ISM in the nucleus then becomes the active ingredient 
for luminosity generation -- via enormously elevated rates of star formation or by feeding a pre-existing massive black hole.  
There new evidence for massive gas and dust accretion disks in a number of these galactic nuclei. Here, we will concentrate on those structures.  
The best example is that found in the ultraluminous infrared galaxy Arp 220. For this object there now exists sub-arcsecond resolution mm-wave aperture synthesis which probes both the molecular gas distribution and the kinematics -- hence the mass distribution (Scoville, 
Yun and Bryant, 1997, Downs and Solomon, 1998, and Sakamoto {\it et al} 1998).  Similar, 
less massive structures are also evident in millimeter-wave  observations of a number of nearby lower level AGN (Tacconi {\it et al} 1998, Baker and Scoville, 1998).

\section{Arp 220}

Arp 220 (IC 4553/4), with an infrared luminosity of 1.5 $\times$ 10$^{12}$~L$_{\odot}$ at $\lambda =$ 8-1000$\mu$m, is one of the nearest ultraluminous 
infrared galaxies (Soifer {\it et al}, 1987).  Visual wavelength images reveal two faint tidal tails, indicating a recent tidal interaction (cf. Joseph and Wright, 1985), and high resolution ground-based radio and near-infrared imaging show a double nucleus (Baan {\it et al}, 1987, Graham {\it et al}, 1990).  The radio nuclei are separated by 0.$^{\prime\prime}$98 at P.A. $\sim$90$^{\circ}$ (Baan and Haschick, 1995). To power the energy output seen in the infrared by young stars requires a star formation rate of $\sim$10$^2$~M$_{\odot}$ yr$^{-1}$.  Alternatively, if the luminosity originates from an active galactic nucleus, this source must be sufficiently obscured by dust that even the 
mid-infrared emission lines are highly extincted since spectroscopy with the ISO shows no evidence of very 
high ionization lines at wavelengths out to 40$\mu$m (Sturm {\it et al}, 1996).

\vskip 0.3cm
\begin{figure}
\includegraphics{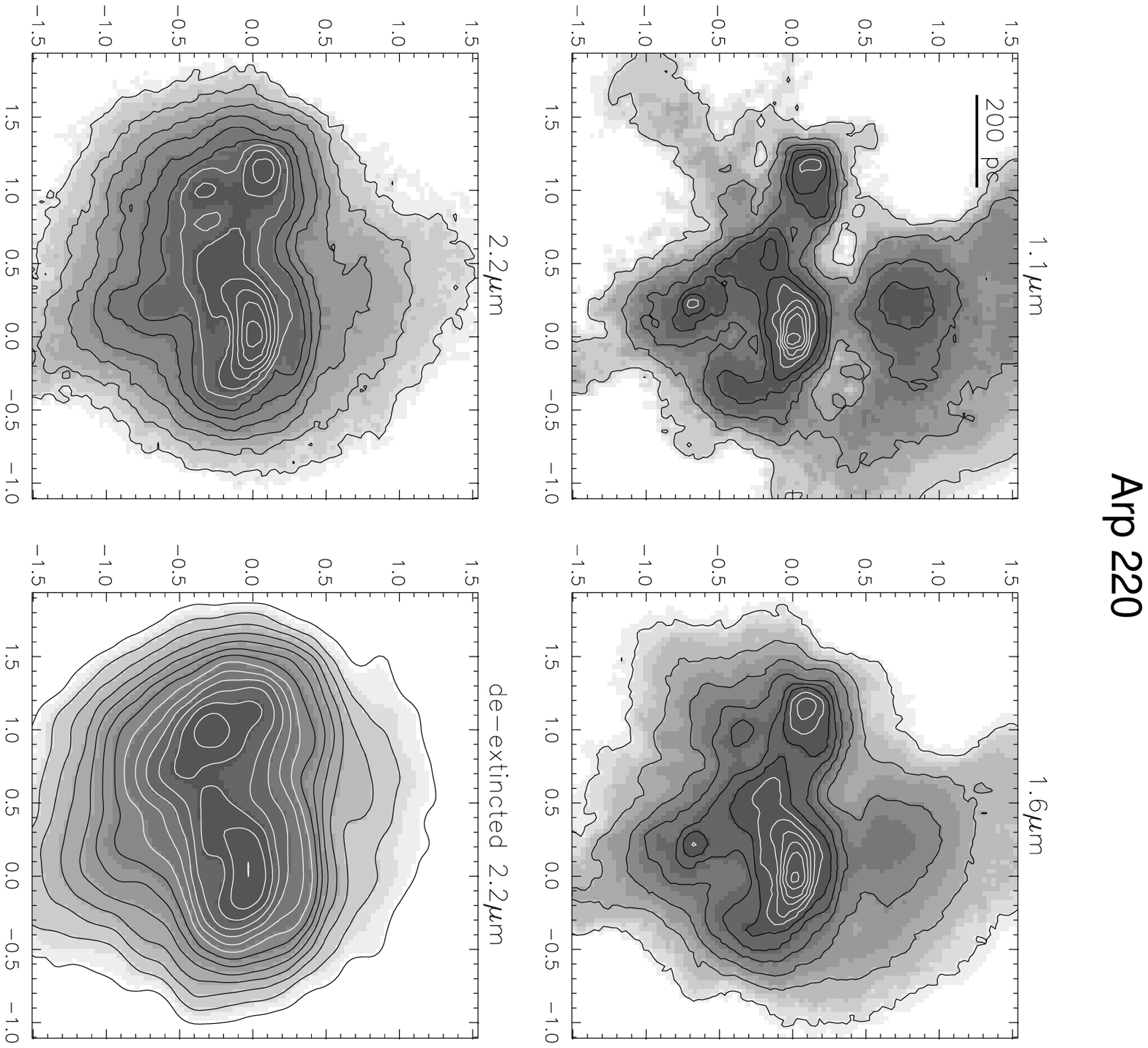}
\vskip 12.0cm

Fig. 1:  Central region of Arp 220 imaged at 1.1 (a), 1.6 (b) 
and 2.2 (c)$\mu$m with NICMOS
on HST (Scoville {\it et al}, 1998).  The lower right panel shows the 'de-extincted' 2.2 $\mu$m image
where the extinction at 2.2$\mu$m was determined from the 1.6/2.2 $\mu$m
color and the assumption of a foreground dust screen. (The lower right
panel has 0.2$^{\prime\prime}$ resolution since both the 1.6 and 2.2 $\mu$m images were smoothed to this resolution to obtain the color). Contours are spaced logarithmically and the axes are labelled in arcsec
offset from the 2.2 (c)$\mu$m peak. The  bar indicates 200 pc. 
\end{figure}

\subsection{Near Infrared}
The central region of Arp 220 is shown in contour form in Figure 1.  Coordinate offsets are measured from the position of peak flux at 2.2$\mu$m occuring on the western nucleus.  These images clearly show the 
two nuclear regions plus several lesser peaks. The morphology of both nuclei changes remarkably with wavelength.  In particular, the bright western nucleus shows greater extension to the
south at 2.2$\mu$m and the eastern nucleus has a southern component which becomes increasingly strong at the longer wavelengths. Under the assumption that the background starlight has an intrinsic color which is not strongly varying, and that the dust producing the extinction is in the foreground, 
ie, not mixed with 
the stars, we have corrected the 2.2$\mu$m image for extinction and use this extinction-corrected image (shown in the lower right panel of Figure 1) to place the near-infrared maps relative to the three cm-wave radio continuum sources in Arp 220.  Our registration places  
one radio nucleus between the two emission peaks seen at 2.2$\mu$m in the east and the western radio nucleus in the area of 
extremely high obscuration to the south of the western 2.2$\mu$m peak.  With this registration, the third, very weak radio component coincides with the faint southern peak seen in the images at all three near-infrared wavelengths (cf. Scoville {\it et al} 1998).

The complex near-infrared structure of Arp 220 is undoubtedly due to there being multiple centers of star formation activity and strongly varying dust obscuration within the merger nuclei.  The crescent or partial ring 
morphology of the western nucleus might readily arise under two circumstances:  if there is an obscuring disk of dust and gas embedded in a spheroidal 
nuclear star cluster, or if a central starburst ring is partially obscured by its own dust on one side.  In either case, the dust (and dense interstellar 
gas) must be confined to a thin disk-like structure, concentric with the western nucleus of the galaxy.  A similar structure probably exists in the eastern nucleus based on the millimeter-wave interferometry (see below) although its inclination may not be so close to 
the line of sight (in view of the less sharp cutoff in the near-infrared light distribution).  The suggested geometry for the western nucleus is a galactic nucleus star cluster cut by an embedded, opaque dust disk at inclination 20$^{\circ}$ to the line of sight.  

\subsection{Molecular Gas}
The total molecular gas content for Arp 220 is 9 $\times$ 10$^9$M$_{\odot}$ based on the CO (2-1) emission and a CO-to-H$_2$ conversion ratio which is 0.45 times the Galactic value (cf. Scoville, Yun and Bryant, 1997). 
This enormous mass (approximately four times that of the entire Galaxy) is contained entirely within R $<$1.5 kpc and approximately 5 $\times$ 10$^9$M$_{\odot}$ is apparently concentrated in a thin disk in 
the nuclear region at radii $<$~250 pc.  
Within the last year there have been two studies of the CO (2-1) emission at 0.5$^{\prime\prime}$ (175 pc) resolution 
(Downs and Solomon, 1998, and Sakamoto {\it et al}, 1998).  In Figure 2, the CO emission and 1.3 mm continuum emission are shown from Sakamoto {\it et al} (1998).  Figures 2a-c show the CO line emission, mean velocity field and 
continuum emission obtained from only the high resolution interferometry data -- that is, filtering out any extended structure by not including short interferometer baselines. Figure 2d shows the total integrated CO emission and mean velocity contours, including both high and low resolution data.  The crosses on 
all four panels indicate the positions of the double radio nuclei (Norris, 1985).  In both the line (a) and the continuum (c), there are clear peaks at the positions of the radio nuclei. 

\vskip 0.3cm
\begin{figure}
\includegraphics{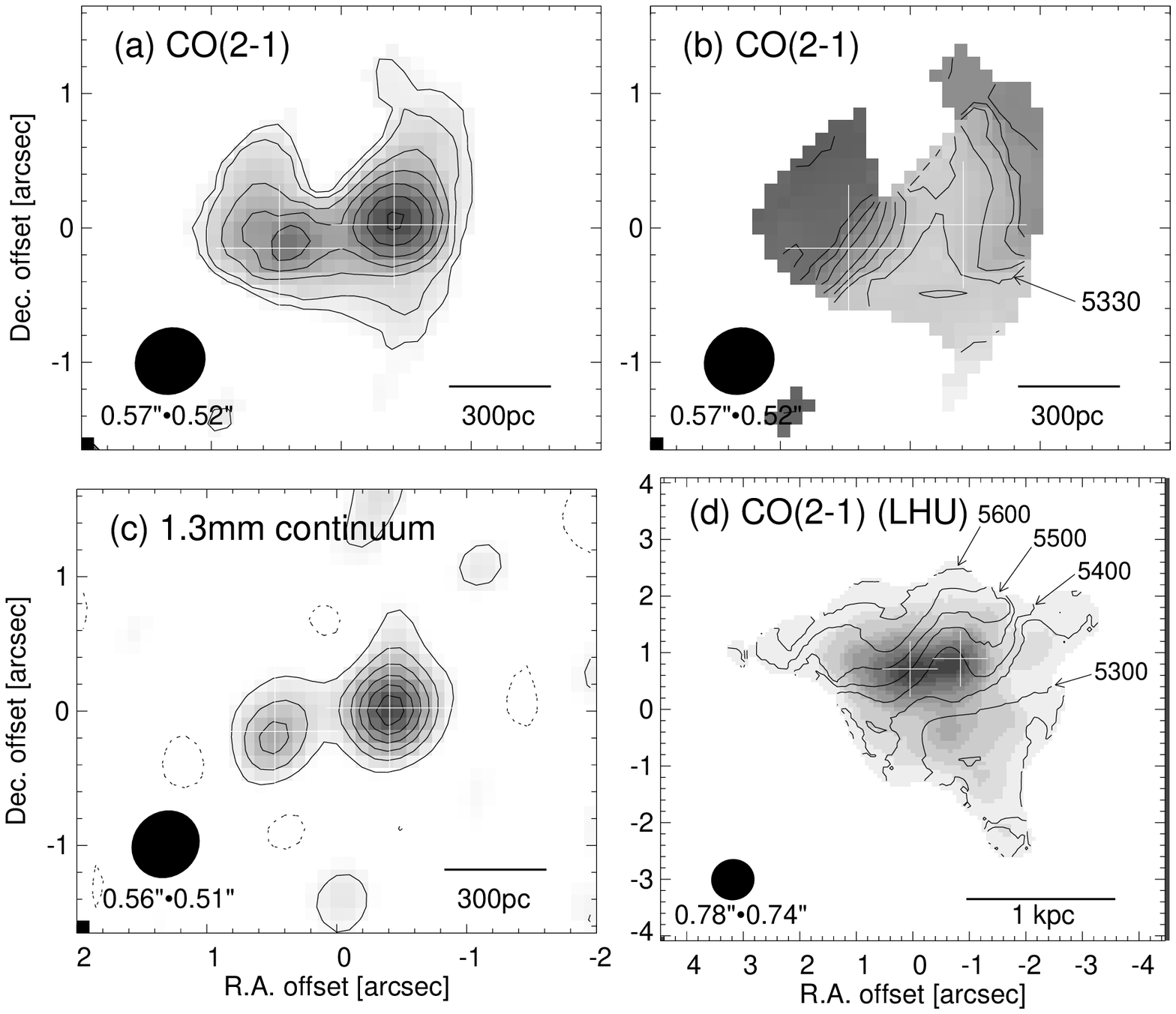}
\vskip 10.5cm

Fig. 2:  Central region of Arp 220.  Crosses in each panel indicate the 1.3 m continuum position of the nuclei. (a) High resolution CO (2-1) emission. (b) Mean velocity of CO (2-1) with contours spaced 50 km s$^{-1}$. (c) 1.3 mm continuum. (d) CO (2-1) integrated intensity map and isovelocity contours (50 km s$^{-1}$ step) made from low(L), high(H), and ultra-high(U) resolution data. 
\end{figure}

The CO peaks coincide with those in the continuum but the line emission is much more extended than the continuum emission.  The diameter of the CO in Figure 2d, in which most of single-disk flux is recovered, is about 2 kpc, and the overall emission (2d) is elongated along a direction
parallel to the gradient in the velocity field.  
The latter provides the basis for modelling the overall emission as a disk since an inclined disk will have parallel kinematic and isophotal major axes (see also Scoville, 
Yun and Bryant, 1997).  The gas disk is, however, not axisymmetric and its velocity field shows distortions, 
which suggest non-circular motions 
or unrelaxed gas components,  probably due to the fact that the system is still an ongoing merger and has two massive nuclear star clusters embedded in the 
disk.

About 30\% of the total CO emission is associated with the two nuclei, and a steep velocity gradient is evident across each nucleus in the high resolution velocity map (Figure 2b).  The velocity shift is about 500 km s$^{-1}$ within 0.$^{\prime\prime}$3 (110 pc) in position-velocity diagrams through each nucleus. 
The velocity gradients across the two nuclei are in opposite directions and neither is aligned with that in the outer molecular disk at P.A. = 25$^{\circ}$ (compare Figures 2b and d).  The very steep velocity gradients at the two nuclei, as well as their directions, exclude the possibility that the gas is just rotating around the dynamical center of the merger.  Instead, the data suggest that each nucleus has a separate rotating molecular gas disk.  
These two disks are counterrotating, and their diameters are 
$\sim$200 pc (0.5$^{\prime\prime}$).  The dynamical mass of each nucleus is $\sim$2 $\times$ 10$^9$ sin$^{2}i$~M$_{\odot}$, where $i$ is the inclination of the disks.  The presence of the two dynamically relaxed disks strongly supports the view that the two peaks (in the millimeter, radio and NIR) are true nuclei, and 
that Arp 220 is indeed a merger.  The major axis of the disk inferred from the NICMOS images is consistent with the E-W velocity gradient at the west nucleus.

The two nuclear disks are embedded in an outer gas disk of kiloparsec radius.  The mean velocity of the two nuclei is different by about 200 km s$^{-1}$, with the eastern nucleus more redshifted.  This is consistent 
with NIR spectroscopy of Br$\gamma$ (Larkin {\it et al}, 1995).  The dynamical mass within the orbit of the two nuclei (ie, $R <$~250 pc) was estimated to be a 5.4 $\times$ 10$^9 $M$_{\odot}$ by Scoville, Yun and Bryant (1997).  Thus more than half of the mass in this region belongs to the two nuclei.  
Note, however, that the dynamical mass of the central disk has large uncertainty due to its poorly constrained inclination (i$\sim45\deg$).

\section{Remarks}

In the ultraluminous galaxies the nuclear 
gas and dust accretion disks are spectacular:  
containing up to 10$^9$~M$_{\odot}$ of ISM at radii 100-200 pc but extremely thin (10-50 pc).  The mean gas densities within the disks are typically 10$^4$ cm$^{-3}$ and, in contrast to our Galactic center where the area-filling factor is only a few percent, 
the disk area is substantially filled (area-filling factors, 20-50\%).
Our results 
show that part of the concentrated gas (30\% in CO luminosity in the case of Arp 220) remains intact around the stellar nuclei to the late stages of merger evolution -- at least until the two 
nuclei approach within several hundred parsecs.  The rest of the gas is rotating around the dynamical center of the merger system. 

It has been shown in numerical simulations that gas rapidly concentrates to the nuclei of merging galaxies at an early phase of interaction.  Subsequently, the concentrated gas quickly merges as the nuclei approach each other (Barnes and Hernquist, 1991). There is strong consensus that the activity in the ultraluminous IR galaxies is triggered by strong dynamical perturbations during a close galactic encounter or galactic merging.  The resultant disruption of the normally circular orbits in a disk system 
will lead to extremely rapid dissipation of rotational energy and angular momentum within the interstellar medium.  Once in the nucleus, any vertical (z) motions of the gas are rapidly damped and dissipated.  The resultant thin disk in 
the nuclear region, containing a galactic mass of interstellar medium, will be vertically supported by turbulence and the radiation pressure of young stars formed in this high density medium.  In fact, such disks may become self-regulating since the stars which form within the disk generate both turbulence and a high radiation pressure. At the elevated rates of star formation seen in these galactic nuclei, the resulting 
turbulent and radiative pressure can swell the disk, and thus damp the star formation.  

The final evolution of 
such disks to AGNs could proceed either along the lines suggested by Norman and Scoville (1988) in which mass-loss from late stellar evolution of the starburst population ultimately accretes into and builds up a super-massive black hole, or 
with a pre-existing black hole being fed directly from the nuclear disk by accretion to very small radii.  In either case, one should not expect the AGN to become 
immediately visible, given the extraordinarily large extinctions (A$_V\geq 10^3$ mag) in the nuclear disk during these early phases. The discovery that the ISM in the late merging systems is often concentrated in a thin, rotationally-supported disk
(rather than in a more chaotic spherical distribution) should enable theoretical 
studies of their evolution to proceed from a more clearly defined initial state.

\acknow
It is a pleasure to acknowledge stimulating discussions with A. Baker, 
C. Norman and K. Sakamoto on this work and assistance with the manuscript
from N. Candelin and Z. Turgel.

\newpage

\newpage

\begin{figure}
\vskip -3.0cm
\hskip 3.0cm
\vskip -0.5cm

Fig. 2:  Central region of Arp 220.  Crosses in each panel indicate the 1.3~mm continuum position of the nuclei. (a) Continuum subtracted CO (2-1) emission integrated over 645 km s$^{-}1$. Contours are at 8.5 $\times$ [1,2,4,6,8,10,12] Jy beam$^{-1}$ km s$^{-1}$. (b) Mean velocity of CO (2-1). Contour interval is 50 km s$^{-1}$. (c) 1.3 mm continuum with contour interval of 18.5 mJy beam$^{-1}$ (2$\sigma$), zero contours omitted, and negative contours dashed. (d) CO (2-1) integrated intensity map and isovelocity contours (50 km s$^{-1}$ step) made from L, H, and U configuration data. Peak integrated intensity is 193 Jy beam$^{-1}$ km s$^{-1}$.

\vskip -1cm
\end{figure}

\end{document}